# Machine learning plastic deformation of crystals

Henri Salmenjoki[1], Mikko J. Alava[1] & Lasse Laurson  [1,2]

Plastic deformation of micron-scale crystalline solids exhibits stress-strain curves with significant sample-to-sample variations. It is a pertinent question if this variability is purely random or to some extent predictable. Here we show, by employing machine learning techniques such as regression neural networks and support vector machines that deformation predictability evolves with strain and crystal size. Using data from discrete dislocations dynamics simulations, the machine learning models are trained to infer the mapping from features of the pre-existing dislocation configuration to the stress-strain curves. The predictability vs strain relation is non-monotonic and exhibits a system size effect: larger systems are more predictable. Stochastic deformation avalanches give rise to fundamental limits of deformation predictability for intermediate strains. However, the large-strain deformation dynamics of the samples can be predicted surprisingly well.

[1] Department of Applied Physics, Aalto University, P.O. Box 11100, FI-00076 Aalto, Espoo, Finland. [2] Laboratory of Physics, Tampere University of Technology, P.O. Box 692, FI-33101 Tampere, Finland. Correspondence and requests for materials should be addressed to L.L. (email: lasse.laurson@tut.fi)

Predicting the behavior of complex, non-linear systems is one of the main challenges of science. Yet, achieving such goals has remained elusive: attempts to forecast phenomena such as earthquakes have so far yielded controversial results at best[1]. Reasonably accurate prediction of, for instance, the failure time of a solid sample subject to external loads[2] or the time of a volcanic eruption[3] has been achieved mostly using information available only relatively close in time to the event one is trying to predict. What makes such predictions difficult is that the mapping from various features describing the state of the system to its subsequent behavior tends to be very complicated. This is due to the non-linear nature of the collective dynamics underlying the time-evolution of the system, as well as the high dimensionality of the feature set characterizing the system's state. This implies that traditional forecasting methods are typically not able to cope with the ensuing complexity.

Recently, the use of artificial intelligence in general and machine learning (ML) in particular have found novel applications in diverse fields and problems including, for instance, image recognition[4], medical diagnosis[5], and statistical arbitrage in finance[6]. A specific variant of machine learning models, artificial neural networks (ANNs), has proven to be particularly useful in discovering meaningful structure in data. Given sufficient amounts of training data, such models, or regression neural networks, are capable of learning complex, non-linear mappings from a high-dimensional feature vector to a desired output. This property makes these models useful to solve novel kinds of problems also in fields such as physics and materials science[7–13], and related activities have very recently gained significant momentum[14].

It is experimentally well-established that micron-scale crystals deform plastically via a sequence of broadly distributed strain bursts, directly visible as steps in the staircase-like stress-strain curve[15–19]. The apparently stochastic nature of the deformation bursts—typically characterized by power-law-like size distributions—results in significant sample-to-sample variability of the stress-strain response. On the other hand, the dynamics of dislocations—the topological defects of the crystal lattice the motion of which mediates the plastic deformation process—should be largely deterministic: in the first approximation, their motion obeys a deterministic mobility law relating the Peach–Koehler force to the instantaneous dislocation velocity[20]. Thus, the details of the deformation process of a given sample are in principle encoded in the features of the initial state, i.e., the pre-existing dislocation network within the crystal. Given a complete characterization of the initial dislocation configuration, the dynamics can be solved from the deterministic equations of motion of the dislocations. The key issues then become to what extent more coarse-grained descriptors of the initial states are sufficient to predict the subsequent bursty deformation dynamics, and what is the role of the apparently stochastic strain bursts on deformation predictability.

Here, we study deformation predictability by applying ML methods able to learn the mapping from features of the pre-existing dislocation microstructure to the ensuing stress-strain curves, using 2D discrete dislocation dynamics (DDD) simulations as a test system. In short, our results show that using a number of physically motivated features to describe the initial states gives rise to robust predictability of the deformation process, the degree of which depends on both the strain level and the system size. We analyze the reasons behind these dependencies, and find that they originate on one hand from the properties of the largely stochastic deformation bursts, and on the other from the predictive power of various descriptors of the initial state evolving with the system size. We also discuss how our results open the door to experimental work on deformation predictability and optimization of mechanical properties of materials.

## Results

**Machine learning plastic deformation**. To study deformation predictability in a simple dislocation system, we start by generating an extensive database of stress-strain curves and the corresponding initial dislocation configurations from 2D DDD simulations (Methods); such a simple model is known to be able to capture the statistics of the strain bursts and hence the fluctuating character of the stress-strain curve[21]. We then train a regression neural network, as well as a support vector machine (Supplementary Note 1 and Supplementary Fig. 1), to infer the mapping from features characterizing the initial dislocation microstructure to the ensuing stress-strain curve. We then study the predictive ability of the model as a function of strain $\varepsilon$ and the system size. For each realization of the simulation, a randomly generated configuration of $N$ dislocations with a zero net Burgers vector (a topological charge of the dislocations) is let to relax in zero external stress, after which the external stress is increased quasistatically (Methods). During this ramp-up of the applied stress, the stress-strain curve describing the deformation process is recorded, see Fig. 1 for examples; in what follows, this is labeled as the basic scenario. To address the role of the preparation of the initial state on deformation predictability, we also perform a set of simulations where the initial states have been obtained by first deforming the basic samples up to a pre-determined strain of $\varepsilon_{ID} = 0.2$ (also intermediate pre-strain values were studied), followed by another relaxation in zero stress; these are referred to as ID (initial deformation) systems. The training, validation and test data sets then consist of a number of features characterizing the initial states with (ID) or without (basic) pre-deformation, as well as of the unique stress-strain curve for each sample. We have tested a large number of physically motivated features. The most important ones in terms of predictive power turn out to be statistical measures of the internal stress field and especially densities of geometrically necessary dislocations (GND)[22] (Methods).

**Deformation predictability**. Once trained, we proceed to analyze the predictive ability of the regression neural network as a function of strain $\varepsilon$, using the standard methodology of evaluating the trained network on a test data set[23] (see also Supplementary Note 1). The performance of the neural network may be quantified by considering the score $S$, defined as $S = 1 - \left[ \sum_i (d_i - y_i)^2 \right] / \left[ \sum_i (d_i - \langle d_i \rangle)^2 \right]$, where $d_i$ and $y_i$ are the desired output (i.e., the stress at certain strain) and network output corresponding to the $i$th test system, respectively; $S = 1$ would correspond to a perfect fit, and the smaller the $S$-value the worse the fit. Notice that using the average stress-strain curve as a benchmark prediction would correspond to $S = 0$. Figure 2a, b show $S$ as a function of $\varepsilon$ considering the basic and ID scenarios, with the stress $\sigma_{ID}$ needed to reach a fixed pre-deformation strain as an additional feature in the latter case. The data shown in Fig. 2 reveals three main observations: first, we notice that the predictive ability $S$ of the ANN is generally larger for the ID samples. For instance, considering the largest system size $N = 400$, $S(\varepsilon = 0.1)$ increases from 0.44 in the case of the basic systems to 0.69 for ID systems. The second key result to notice is that $S$ exhibits a clear non-monotonic dependence on $\varepsilon$, such that it initially decreases up to a characteristic strain $\varepsilon^*$ in the range between 0.02–0.03. Surprisingly, for larger $\varepsilon$, $S$ starts to increase again. As the final observation, the large-strain predictability displays a non-trivial size effect. In the basic scenario, the effect is monotonic and the score of the largest system is far better than the corresponding score of the smallest system. The largest ID systems have the largest $S$, but for smaller $N$ the size dependency becomes non-monotonic. Considering a support vector machine instead of ANN leads to almost identical $S(\varepsilon)$ curves, verifying the general nature of our results (Supplementary Fig. 2).

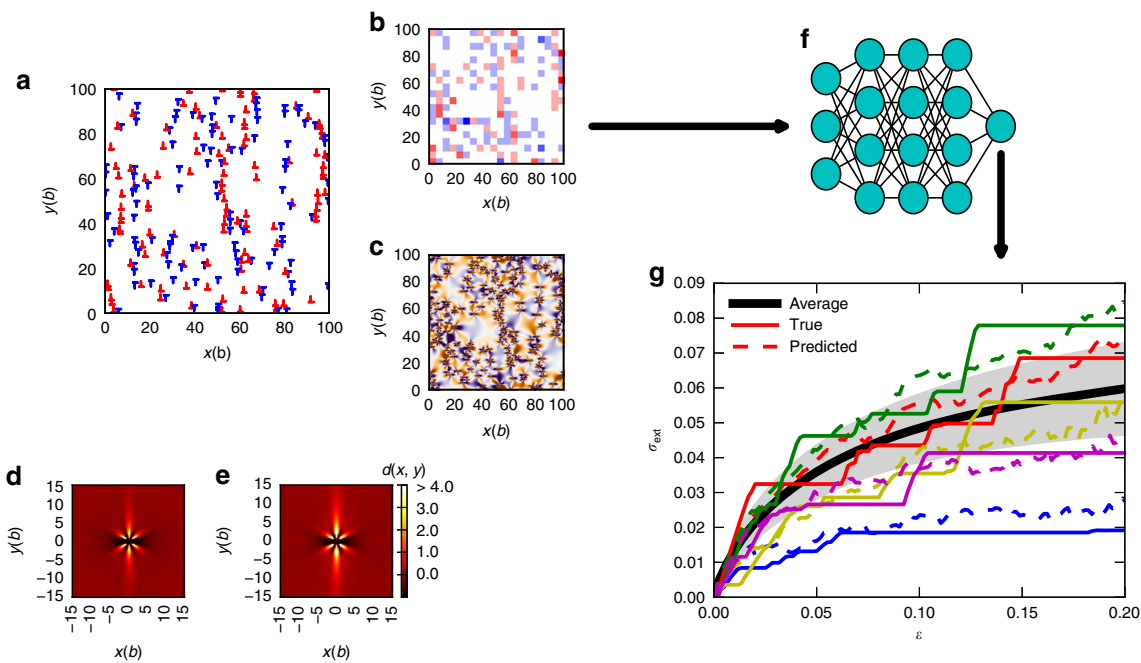

**Fig. 1** Machine learning plastic deformation. Using features of the initial dislocation configuration (**a**) (with red and blue symbols corresponding to positive and negative Burgers vectors, respectively), such as the density of geometrically necessary dislocations (depicted in (**b**), with red and blue corresponding to positive and negative values of $\rho_{GND}$, respectively) or the internal stress field (in (**c**), orange corresponding to positive and purple to negative $\sigma_{sf}$), and considering as initial states relaxed random configurations (basic, with the dislocation pair correlation function in (**d**)), as well as pre-deformed dislocation configurations (ID, (**e**)), we train a neural network (a schematic is shown in (**f**)) to infer the relation between features of the initial dislocation configurations and the ensuing stress-strain curves. Examples of the true (solid lines) and predicted (dashed lines) curves for a few basic samples, along with the average (thick black line) and standard deviation (shaded region) of the true stress-strain curves are shown in (**g**). All the figures are from systems with 400 initial dislocations

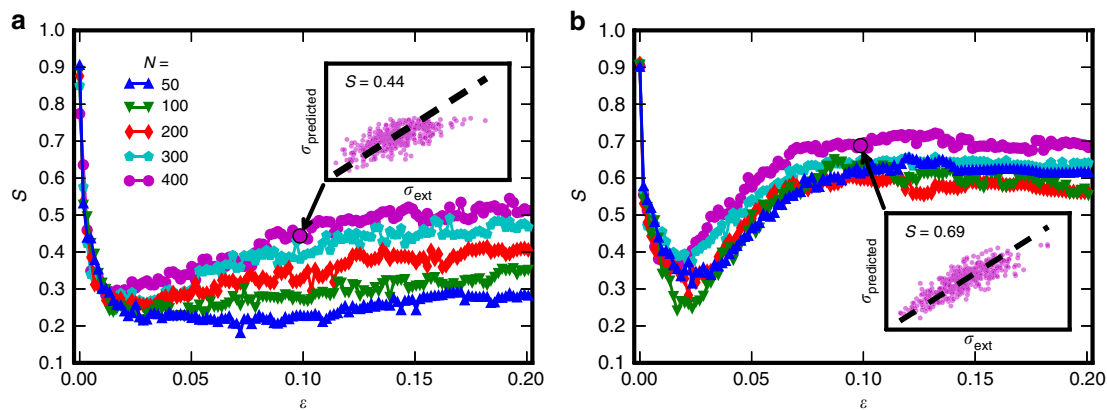

**Fig. 2** Predictability of the stress-strain curves: large systems are more predictable for larger strains. Score $S$ of the ANN prediction as a function of strain $\varepsilon$ for basic (no pre-strain, (**a**)) and ID (with a pre-strain of 0.2, (**b**)) systems of different sizes. The insets show example scatter plots of the predicted vs simulated stress values for $N = 400$ systems at $\varepsilon = 0.1$

**Pre-deformation improves deformation predictability**. Starting from the difference between the predictability of the basic and ID systems, we attribute this observation to the fact that pre-deformation leads to formation of dislocation structures, resulting in more clear-cut features and hence better predictability. Predictability of ID systems is better especially for large strains ($\varepsilon > \varepsilon^*$), with the increase of $S$ with $\varepsilon$ for $\varepsilon > \varepsilon^*$ significantly more pronounced than in the basic case. The difference in the initial structures is evident in the average pair correlation function $d(x, y) = \langle \sum_i \rho_i(x, y)/\rho_0 \rangle - 1$, where $\rho_0$ is the average dislocation density and $\rho_i(x, y)$ is the dislocation density at position $(x, y)$ relative to dislocation i (Fig. 1d, e). $d(x, y)$ shows stronger correlation characteristic to these systems in the ID case: dislocation walls and dipoles are more noticeable[24].

Figure 3 emphasizes this in the case of the largest system considered ($N = 400$), by illustrating how the dislocation pair correlations decay along the dislocation walls ($y$-direction) for systems with different amounts of pre-strain. The correlation function $d(0, y)$ of all dislocations is shown in Fig. 3a, while the correlation function of positive dislocations $d_{++}(0, y)$ is displayed in Fig. 3b. The general observation is that the larger the pre-strain, the stronger the correlations, and correspondingly, the more correlated systems become more predictable towards the end of the simulation. We quantify this by power-law fits of the form of $d(0, y) = \alpha y^{-\beta}$. For the correlation function $d(0, y)$ of all dislocations, we find a pre-strain dependent power-law exponent $\beta$, see Fig. 3a. The inset of Fig. 3a shows that the average prediction score in the strain range $\varepsilon \in [0.15, 0.2]$ decreases with increasing $\beta$ (i.e., faster decay of the correlation function). The corresponding correlation functions of positive dislocations, $d_{++}(0, y)$, exhibit a power-law decay with an exponent $\beta$ close to 1.5, in agreement with previous results[25,26] (see Fig. 3b). In this case the amplitude $\alpha$ of the power-law fit $d_{++}(0, y) = \alpha y^{-\beta}$ increases with the imposed pre-strain, and can be used as an alternative way to quantify correlations. The inset of Fig. 3b shows that large-strain predictability improves with increasing $\alpha$, and hence with the correlations. We also note that the exponent $\beta$ of the correlation function $d(0, y)$ of all dislocations appears to approach the value 1.5 from above as the amount of pre-strain is increased, possibly due to the wall-like structures within the system being increasingly composed of dislocations of the same sign. Interestingly, the observed effect of the pre-strain on predictability is monotonic only in the large strain regime (Supplementary Figs. 3 and 4). This arises due to two competing phenomena, namely the dislocation structures introduced by the initial deformation and the non-monotonic strain dependence discussed below.

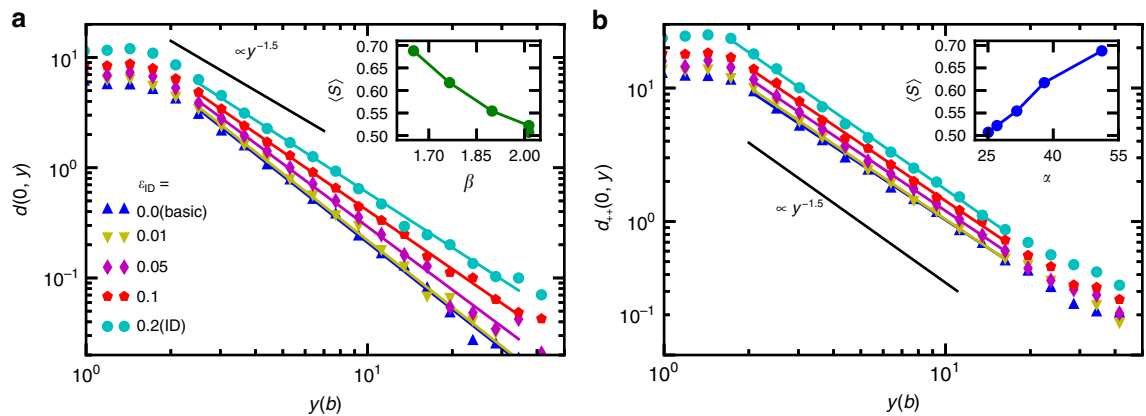

**Fig. 3** Dislocation correlations and predictability. **a** Average pair correlation functions $d(0, y)$ of initial dislocation configurations along the $y$-direction for $N = 400$ obtained with varying amount of pre-strain (See Fig. 1d,e for the full $d(x, y)$ in basic and ID systems, respectively). With larger pre-strain the correlations, i.e., dislocation walls, become larger. The colored solid lines show fits of the form of $d(0, y) = \alpha y^{-\beta}$ and the black solid line illustrates $\alpha y^{-1.5}$ for reference[25,26]. The inset shows that smaller $\beta$ implies better predictability with large strains as the average score in the range $\varepsilon \in [0.15, 0.2]$ (for full curves see Supplementary Fig. 3) decreases with growing $\beta$. **b** The correlation functions $d_{++}(0, y)$ of positive dislocations with similar power-law fits. Now $\beta$ is close to 1.5 so the pre-factor $\alpha$ versus the average score is plotted in the inset

**Predictability and avalanche activity.** We then proceed to analyze the reason behind the surprising non-monotonic strain dependence of deformation predictability. It is well-known that the deformation process of small-scale crystalline systems consists of a sequence of strain bursts with a power-law-like size distribution[15–17,21]. It is often argued that systems with such scale-free dynamics operate in the proximity of a critical point of a non-equilibrium phase transition[27], or within an extended critical region spanned by a range of control parameter values[21]. If deformation bursts are indeed critical avalanches in analogy to those in, say, Barkhausen noise[28], the occurrence of a deformation burst of a given size at a specific stress should be intrinsically hard to predict: In the absence of complete characterization of the system, they may be described as uncorrelated random variables with a probability distribution.

The deterioration of deformation predictability observed in Fig. 2 seems to be due to the onset of large deformation bursts, which are not easily predictable using the coarse-grained description of the system considered here; this is also seen in

Fig. 1g where the predicted stress-strain curves reproduce the typical shapes but not the individual strain bursts of the real curves. Sometimes the predicted curves in Fig. 1g even have short segments where the stress decreases with strain, but this just reflects the fact that the ML algorithm has not learned the stress-strain curves perfectly (stress is never a decreasing function of strain in the simulated curves due to the loading protocol employed). Figure 4 shows the probability distribution function (PDF) of $\varepsilon_{aval}$, the starting strain of a deformation burst, along with the score curves now plotted with a logarithmic strain axis. In both scenarios, basic in Fig. 4a and ID in Fig. 4b, the large decrease of the score until $\varepsilon^*$ coincides with the onset of avalanche activity and the score minima are aligned with the distribution maxima (which is seen also when comparing the scores and $\varepsilon_{aval}$ PDFs of systems with different amounts of pre-strain, see Supplementary Fig. 4). This is related also to the recently investigated concept of first pop-in or discrete plastic event[29], as well as to the deformation avalanches in the microplastic regime recently proposed to be governed by weakest link arguments[30].

$S$ plotted on a logarithmic strain scale reveals that there is an additional local minimum in the predictability around $\varepsilon = \varepsilon^{**} \approx 3 \cdot 10^{-4}$. This minimum is not as dramatic as the one at $\varepsilon^*$ and it originates most likely from the nature of the used feature set: With $\varepsilon < \varepsilon^{**}$ the systems are still almost identical to the relaxed initial states where the interaction between dislocations is insignificant and the stress response is trivial. Around $\varepsilon \approx \varepsilon^{**}$, the dislocations start to feel the presence of the other dislocations, but due to the coarse scale of used features, the stress response here is hard to predict. When $\varepsilon > \varepsilon^{**}$, dislocation displacement becomes significant, and the coarse-grained features start to contain relevant information. Hence the predictability improves up to the onset of burst activity.

**Size effect in deformation predictability.** Finally we address the system size effect seen in the $S(\varepsilon)$ curves of Fig. 2, i.e., larger systems are more predictable for $\varepsilon > \varepsilon^*$. Plastic deformation of micron-scale samples is in general dependent on the sample size, and well-known size effects such as smaller systems being stronger have been reported[17,31] (this is evident also in our simulations, see Supplementary Fig. 5; notice that in our simulations this size effect arises with periodic boundaries, while in micropillar compression experiments the presence of open boundaries is important). Figure 5a illustrates the PDF of avalanche sizes $s$ with starting strains in a bin relatively close to the start of the simulation with the basic systems (for distributions of all avalanche sizes, see Supplementary Fig. 6). As the simulation advances along the stress-strain curve, the bursts become larger with the distribution cut-offs shifting towards larger burst sizes[21], and the smaller systems tend to always exhibit larger bursts (similar in ID systems, Supplementary Figs. 7 and 8). Thus, $S$ of larger systems is less affected by the strain bursts as they tend to be smaller.

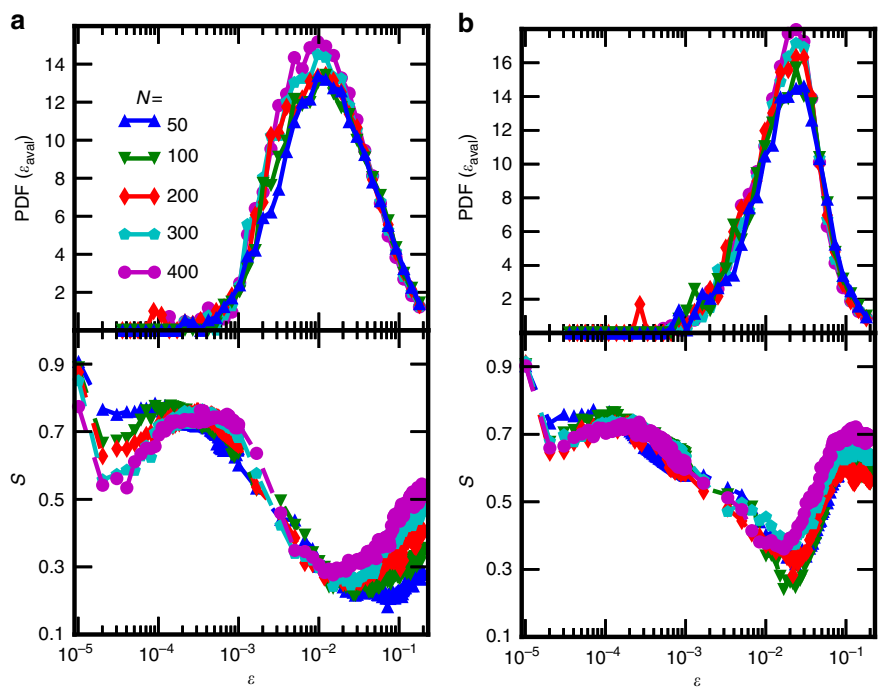

**Fig. 4** Stochastic strain bursts lead to deterioration of deformation predictability. Top: Probability density of the starting strain of the avalanches, $\varepsilon_{aval}$, in **a** basic and **b** ID systems. Bottom: The same ANN score $S$ as in Fig. 2 (with **a** showing data for the basic and **b** for the ID systems) with a logarithmic strain axis, revealing also an additional local minimum of $S(\varepsilon)$ for small $\varepsilon$

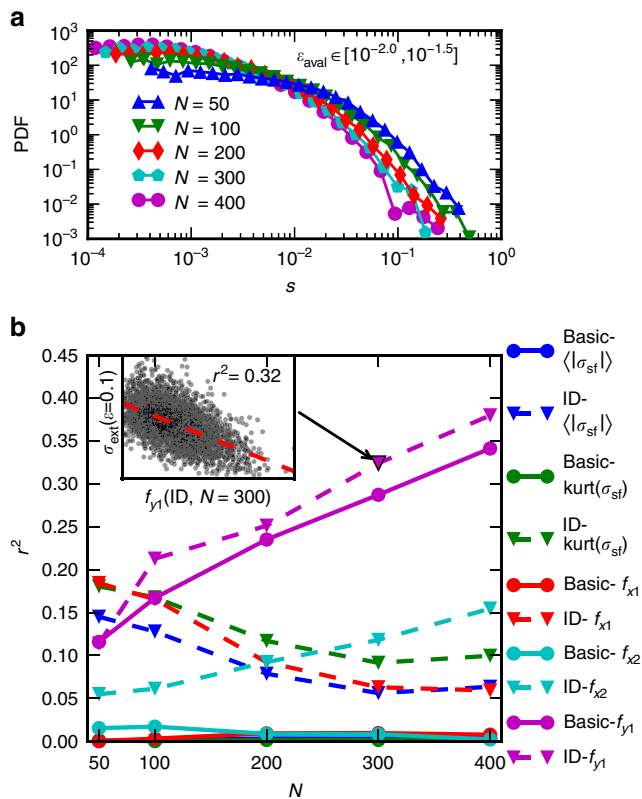

**Fig. 5** Size effect of predictability. **a** The probability density of the strain avalanche sizes $s$ in the basic system for bursts starting in a strain bin close to the start of the simulation. **b** $r^2$ of linear fits between chosen feature parameters (Methods) and the stress $\sigma_{ext}(\varepsilon = 0.1)$ as a function of the system size (inset shows an example of such a fit). Further information of the $r^2$ of single parameter fits for stresses at different strains is presented in Supplementary Fig. 9

**Parameter significance.** Figure 5b shows $r^2$ values of linear fits between $\sigma_{ext}(\varepsilon = 0.1)$ and various input parameters of the ANN as a straightforward measure of parameter significance in systems with different sizes, and partly explains the non-monotonic size effect of the ID scores. In the basic scenario, the size effect implying that larger systems are more predictable is mostly due to the increasing information in the parameter $f_{y1}$ describing the GND density imbalance in the $y$-direction. Meanwhile in the ID scenario, there are parameters related to the internal stress field that are notably significant in addition to $f_{y1}$, but the significance of the stress field descriptors decreases in larger systems. Additionally, the $r^2$ values show that, by itself, $f_{y1}$ is the most informative descriptor. This arises from the fact that it is conserved: it describes the dislocation imbalance in the $y$-direction, and as the dislocations move along their glide planes in the $x$-direction, $f_{y1}$ is constant throughout the simulation. The significance of $f_{y1}$ is even more emphasized with larger strains as other descriptors lose their relevance (see Supplementary Fig. 9). One should note that while the $r^2$ values reveal interesting information about the significance of the various descriptors employed in isolation, ANN and SVM predictions discussed above are non-linear mappings from all the descriptors of the initial states to the stress at a given strain.

## Discussion

To summarize our findings, we observed predictability in the highly fluctuating stress response of a model of a plastically deforming crystal. Since one classical definition of yield strength is the stress corresponding to a given strain threshold, our results can be directly interpreted as predictions of the sample yield strength corresponding to different strain thresholds, with the score parameter $S$ quantifying how well the prediction works for different strains thresholds, pre-strains, and systems sizes. The predictability is significantly better than what one would obtain by using the average stress-strain curve as the prediction (as $S > 0$ for all $\varepsilon$), and exhibits two local minima as a function of the strain; the first and minor one originating from the coarse nature of the used feature set and the second and more dominant one due to the onset of significant avalanche activity. The predictability recovers after the deformation bursts become less frequent and the score of the predictions is surprisingly good towards the end of the simulations. Larger systems are found to be more predictable. This is due to larger strain avalanches present in the smaller systems causing larger aberration to the stress response, as well as due to the fact that the dislocation configurations in larger systems contain more information useful for deformation predictability.

Our study could be generalized to 3D DDD simulations[32], as well as to models containing quenched disorder interacting with the dislocations[27]. In 3D dislocation systems with multi-slip conditions, predictability might be affected by the fact that in that case $f_{y1}$ should not remain constant during straining, unlike in the present case of a 2D model with a single-slip geometry. On the other hand, in 3D under multi-slip conditions forest dislocations on inactive slip systems might provide features which do not evolve much during straining and should thus be useful for prediction. Incorporating quenched disorder to the models—e.g., to mimic precipitate particles[33]—would increase the dimensionality of the relevant feature vector: statistical characteristics of the resulting pinning landscape in a given sample should contain additional information useful for deformation predictability. The pinning points also alter the avalanche statistics as the system starts to exhibit a depinning transition[27].

It would be interesting to test our ideas experimentally considering 3D imaging data obtained by various X-ray measurement techniques[34–40], or optical microscopy of colloidal crystal experiments[41], to construct features of the initial dislocation microstructure, and then try to predict the consequent sample strength for instance at $\varepsilon = 0.1\%$ which is one of the typical definitions for yield strain. Another avenue of future research might be given by application of ML based optimization algorithms (such as Bayesian optimization[42]) to design material microstructures giving rise to samples with desired mechanical properties, such as a large yield stress, or small deformation fluctuations. Our results thus provide novel insight into deformation predictability of materials, and should find applications in fields such as materials design and optimization.

## Methods

**DDD simulations.** To generate the data sets, we consider a 2D DDD model similar to the one in Refs. [21,27], describing a set of parallel, straight edge dislocations with equal number of positive and negative Burgers vectors of magnitude $b$. The dislocations move in a square box of size $L$ with periodic boundaries and interact via the dislocation-generated shear stress fields[20], $\sigma_d(\mathbf{r}) = \sigma_d(x, y) = Dbx(x^2 - y^2)/(x^2 + y^2)^2$, where $D = \mu/2\pi(1 - \nu)$, with $\mu$ and $\nu$ the shear modulus and Poisson ratio, respectively. The overdamped equations of motion describing the glide motion of dislocations along the $x$-direction are $1/(\chi b)v_i = s_i b \left[ \sigma_{ext} + \sum_{i \neq j} s_j \sigma_d(\mathbf{r}_j - \mathbf{r}_i) \right]$, where $\chi$ is the dislocation mobility, $s_i$ and $s_j$ are the signs of the Burgers vectors of dislocations $i$ and $j$, respectively, and $\sigma_{ext}$ is the external stress. We measure lengths in units of $b$, time in units of $1/\chi b D$ and stresses in units of $D$. To mimic dislocation annihilation, two dislocations with opposite Burgers vectors are removed from the system if their distance is less than $2b$. We consider different initial numbers $N$ of dislocations ranging from 50 to 400 distributed randomly within the simulation box, adjusting $L$ so that the dislocation density $\rho = N/L^2$ is kept constant. The initial states of the basic scenario are obtained by relaxing these random configurations with $\sigma_{ext} = 0$. Then, we increase $\sigma_{ext}$ quasistatically from zero: $\sigma_{ext}$ is increased at a slow rate whenever the strain rate is below a small threshold, and is kept constant during the strain bursts (see also Supplementary Fig. 10 and

Supplementary Table 1 for more details on the strain bursts.). The ID initial states are obtained by relaxing systems quasistatically pre-deformed up to a strain $\varepsilon_{ID} = 0.2$ with $\sigma_{ext} = 0$, after which another quasistatic stress ramp is performed.

**Descriptors for machine learning.** The relaxed initial states are characterized by different descriptors. (i) Statistical features of the stress field $\sigma_{sf}$ generated by the dislocations, and its absolute value: Average, median, variance, skewness, and kurtosis. To avoid problems due to diverging stress values if dislocations appear near grid points where the stress field is computed, the limit $|\sigma_{sf}| \leq 2.0$ is imposed (other threshold values yield similar results). Notice that without such a limit, sampling the stress field using a grid of points would sometimes result in arbitrarily large values which would completely dominate the statistical properties (average, etc.) of the stress distribution, an artifact of using linear elasticity. (ii) Density of geometrically necessary dislocations, $\rho_{GND} = \rho_+ - \rho_-$, where $\rho_+$ and $\rho_-$ are the densities of dislocations with positive and negative Burgers vectors, respectively, in slices of width $b$ parallel and perpendicular to the glide planes. Due to the periodic boundaries, we use their Fourier coefficients as features (cross terms did not improve the predictions); thus, coefficients $f_{xi}$ and $f_{yi}$ were included, where $x$ and $y$ refer to the direction of the slicing and $i$ to the $i$th coefficient). (iii) We identified dislocation walls by observing that dislocations with glide planes separated by less than $\sim 10b$ and positions along glide planes separated by less than $\sim 3b$ tend to move collectively, and calculated the number and maximum and average heights of such structures. Finally, the number of dislocations after the relaxation, and also the stress $\sigma_{ID}$ at the end of the pre-shear for ID systems are used as features. All the descriptors considered are collected to Supplementary Table 2. The feature set is then used to train a neural network and a support vector machine to predict $\sigma_{ext}$ as a function of $\varepsilon$, considering data sets of 5000 (10,000 for $N = 50$) samples. Technical details of the model implementation can be found in Supplementary Note 1. For a thorough introduction to the ML methods we refer to ref. [23].

## Data availability

The data that support the findings of this study are available from the corresponding authors on reasonable request.

Received: 8 June 2018 Accepted: 23 November 2018
Published online: 13 December 2018

## Acknowledgements
This work has been supported by the Academy of Finland through an Academy Research Fellowship (LL, project no. 268302). We acknowledge the computational resources provided by the Aalto University School of Science Science-IT project, as well as those provided by CSC (Finland).

## Author contributions
L.L. and M.J.A. designed the study. H.S. performed the numerical simulations and implemented the machine learning algorithms, and contributed significantly to the final form of the study. L.L. wrote the first version of the manuscript. All authors contributed to improve the manuscript.

## Additional information
**Supplementary Information** accompanies this paper at https://doi.org/10.1038/s41467-018-07737-2.

**Competing interests:** The authors declare no competing interests.

**Reprints and permission** information is available online at http://npg.nature.com/reprintsandpermissions/

**Journal peer review information:** *Nature Communications* thanks the anonymous reviewers for their contribution to the peer review of this work.

**Publisher's note:** Springer Nature remains neutral with regard to jurisdictional claims in published maps and institutional affiliations.